# dsLassoCov: a federated machine learning approach incorporating covariate control

Han Cao[#a], Augusto Anguita[#b,h], Charline Warembourg[c], Xavier Escribà-Montagut[b,h], Martine Vrijheid[b,h], Juan R. Gonzalez[b,h], Tim Cadman[b,g], Verena Schneider-Lindner[d], Daniel Durstewitz[a], Xavier Basagaña[b,h*], Emanuel Schwarz[e,f*]

[#] These authors contributed equally to this work

[*] Corresponding authors

[a] Department of Theoretical Neuroscience, Central Institute of Mental Health, Medical Faculty, Heidelberg University, Germany

[b] ISGlobal, Barcelona, Spain.

[c] Univ Rennes, Inserm, EHESP, Irset (Institut de recherche en santé, environnement et travail) - UMR_S 1085, F-35000 Rennes, France.

[d] Department of Anesthesiology and Surgical Intensive Care Medicine, Medical Faculty Mannheim, Heidelberg University, Theodor-Kutzer-Ufer 1-3, 68167, Mannheim, Germany

[e] Hector Institute for Artificial Intelligence in Psychiatry, Central Institute of Mental Health, Medical Faculty Mannheim, Heidelberg University, Mannheim, Germany

[f] Department of Psychiatry and Psychotherapy, Central Institute of Mental Health, Medical Faculty, Heidelberg University, Mannheim, Germany

[g] Genomics Coordination Center, Dept of Genetics, University Medical Centre Groningen, Groningen, Netherland

[h] Universitat Pompeu Fabra (UPF), Barcelona, Spain; CIBER Epidemiologa y Salud Pública, Madrid, Spain.

## Abstract

Machine learning has been widely adopted in biomedical research, fueled by the increasing availability of data. However, integrating datasets across institutions is challenging due to legal restrictions and data governance complexities. Federated learning allows the direct, privacy preserving training of machine learning models using geographically distributed datasets, but faces the challenge of how to appropriately control for covariate effects. The naive implementation of conventional covariate control methods in federated learning scenarios is often impractical due to the substantial communication costs, particularly with high-dimensional data. To address this issue, we introduce dsLassoCov, a machine learning approach designed to control for covariate effects and allow an efficient training in federated learning. In biomedical analysis, this allow the biomarker selection against the confounding effects. Using simulated data, we demonstrate that dsLassoCov can efficiently and effectively manage confounding effects during model training. In our real-world data analysis, we replicated a large-scale Exposome analysis using data from six geographically distinct databases, achieving results consistent with previous studies. By resolving the challenge of covariate control, our proposed approach can accelerate the application of federated learning in large-scale biomedical studies.

**Introduction**

Big data technologies have been widely adopted in biomedical research, with machine learning (ML) analysis demonstrating significant success in various molecular[1] and clinical[2] applications. This success stems from the development of accurate and generalizable models through training on extensive datasets. However, a central challenge in biomedical research is that data sources are often confined to isolated databases within individual institutes or hospitals. These databases often operate under distinct standards and governance structures[3], limiting the ability to pool data effectively. The impediment to data pooling is also made difficult by policy regulations, such as the General Data Protection Regulation[4] (GDPR) in the EU, which requires explicit consent from the patients for the transfer of their data[3]. This challenge significantly hampers the generalizability of ML technologies, e.g., deep learning, which heavily relies on access to extensive training data, and becomes more pronounced in data-scarce areas, such as rare disease analysis[5]. As highlighted in a recent review[6], federated learning emerges as a promising cross-silo ML technology that safely facilitates model learning by leveraging all available data.

Federated learning[7] (FL) is a distributed ML approach designed to enhance data privacy protection. In the context of cross-silo scenarios, the FL model undergoes training on multiple isolated databases, often geographically dispersed. Each database resides behind the firewall on the server of the corresponding institution, and a network with limited access is established among these distributed databases to enable authorized access. Here, only non-disclosive parameters of models[8], devoid of sensitive information, are allowed to be exchanged among the database servers. FL provides the advantage of leveraging multiple datasets for ML model training whilst reducing concerns about the disclosure or exposure of individual-level information. This innovative approach addresses important ethical and legal challenges inherent in data analysis in biomedicine. FL has demonstrated significant potential in various domains, including mitigating data imbalance issues through federated Genome-Wide Association Study (GWAS) analysis[9,10], as well as detecting rare diseases through the federated integration of data cohorts[11].

A critical challenge encountered in FL is addressing the issue of statistical heterogeneity[12]. Statistical heterogeneity refers to the inconsistency of data distributions across different datasets, which violates an assumption underlying most ML methods. For example, typical ML methods assume an identical independent distribution of data (I.I.D.), a condition unlikely to be met in FL applications in general. This discrepancy arises due to a couple of reasons. First, data in different servers are often collected and managed using distinct governance structures, inevitably introducing technical counfounding[13]. Second, since studies are typically situated in geographically diverse locations and tend to collect data from locally recruited subjects, biological counfounding[13], such as effects related to genetic population structure, need to be accounted for. These confounding effects often distort the observed associations between outcomes and variables of interest, complicating interpretation and limiting the reliability of findings[14].

Traditional methods for controlling confounding effects called covariate control approaches, extensively studied in biomedicine. Covariates, referring to confounders and their effects, can be either removed during the preprocessing stage of the ML[15] or controlled throughout the ML process[16]. The former approach, known as the "regressing out covariates" approach, entails regressing out covariates from each feature and outcome. The residuals, considered covariate-free, are then used in subsequent ML analysis. The latter approach, the "regression adjustment" approach, incorporates covariates into the linear model to control for their effects. Both approaches are proven equivalent in a low-dimensional setting, as guaranteed by the Frisch–Waugh–Lovell theorem[17]. However, these approaches are not always feasible in FL applications. The "regressing out covariates" approach is feature-wise, which means that calculations need to be conducted feature by feature, resulting in computational intractability in high-dimensional FL problems. Additionally, it is only suitable for linear regression problems, limiting its applicability to classification problems. The "regression adjustment" approach also requires a low-dimensional setting to avoid overfitting, rendering it impractical for high-dimensional scenarios.

Given the aforementioned considerations, our current paper introduces 'dsLassoCov' — to the best of our knowledge the first federated ML approach capable of controlling confounding effects in high dimensional settings. We implemented and validated dsLassoCov for constructing FL linear models and for feature selection in the presence of confounding effects. Using simulated data, we demonstrate that our approach achieves higher computational efficiency and feature selection accuracy compared to conventional ML workflows designed to address confounding effects. Furthermore, we show that dsLassoCov is mathematically equivalent to an iterative procedure that alternates between a covariate removal step and a parameter update step. This guarantees the estimation of "covariate-free" parameters at each iteration. Finally, we illustrate its practical application through its utilization on multiple real-world exposome federated datasets derived from the HELIX (Human Early-Life Exposome) Project[18].

**Methods**

Our implementation of dsLassoCov, which means dataShield LASSO regression adjusted for covariates, is based on the dsMTL[19] framework and DataSHIELD [8]. In this method section, we first introduce what is the DataSHIELD environment. Then, we provide an overview of the intuition behind dsLassoCov. We outline the objective that dsLassoCov aims to minimize, followed by a description of the optimization process and detailed algorithm. The technical intricacies are condensed and provided in the **supplementary methods** section. Detailed information about DataSHIELD can be found also in the **supplementary methods**. Next, we proceed by elucidating the covariate control mechanism employed by dsLassoCov. Finally, we describe the simulation and real data analyses results.

*DataSHIELD*

DataSHIELD (ds) is a software created for data analysis at an individual level using disclosure-preventing methods that address ethical and legal restrictions concerning confidentiality[20,21]. Principal features of DataSHIELD involve: I) a client-server architecture ("taking the analysis to the data"), II) analytical methods for fedeated analysis including both mega and meta-analyses for the federated scenario, and III) tailored multi-layer disclosure controls (the bottom line being that the analyst cannot see, copy or extract individual level data held by individual studies). The key aspects of DataSHIELD include its open-source nature, that is written in R, and is licensed under the GPL (General Public License), thus facilitating downstream analyses within a single pipeline by interacting with other programming languages (e.g., Python) and with other R or Bioconductor packages. The fundamental building blocks of DataSHIELD are its client-side and server-side functions. Server-side functions reside in the modified R environments located behind the firewall of the data computers at each individual study. It is the server-side functions that actually process the individual-level data at the distinct repositories. The outputs from server-side functions (non-disclosive study-level statistics) represent the only information that ever leaves a data computer, and this is why we can claim that DataSHIELD allows full analysis of individual-level data without those data ever having to be moved, or even rendered visible, outside their study of origin. Client-side functions reside on the conventional R environment on the analysis computer. Client-side functions call and control server-side functions and combine information across different repositories when required. All DataSHIELD functions require approval under a technical and governance process including external independent evaluation.

*Intuition behind dsLassoCov*

Consider a "regression adjustment" approach[16], which effectively controls the confounding effect in both regression and classification problems, but encounters limitations in high-dimensional settings due to the overfitting problem. This weakness can be mitigated by integrating the LASSO (Least Absolute Shrinkage and Selection Operator)approach[22], which has demonstrated robustness against overfitting. LASSO is a regularization technique among the most straightforward and frequently employed for feature selection[23]. The LASSO regression is similar to ordinary least squares, except that it imposes a shrinkage penalty on the magnitude of the estimated coefficients[22,24]. In the current paper, we present dsLassoCov, a federated version of LASSO with the ability to controls for confounders in both regression and classification problems. As shown in **Figure 1**, the regularization path of dsLassoCov is illustrated, showing the outputs of model training and creating a series of models indexed by a hyperparameter λ[23]. A larger λ value corresponds to a model with fewer selected features, as illustrated in **Figure 1**. Therefore, integrating the principle of LASSO can create a low-dimensional environment by tuning λ, which enables the successful application of covariates removal methods. For instance, in **Figure 1**, when λ=0.07, the corresponding model contains three features, and the "regression adjustment" approach can be safely applied.

In the subsequent sections, we will provide a mathematical interpretation of our method concerning its ability to control covariates. We will demonstrate that dsLassoCov follows an iterative procedure, wherein covariates removal is repeatedly performed alongside parameter estimation in each iteration. Additionally, we will illustrate that dsLassoCov allows for the utilization of more complex models while maintaining its covariate control capability.

*Modeling, optimization and algorithm*

$$\min_{w,w_c} \frac{1}{2n} \sum_{i=1}^{n} (y_i - x_i w - x_i^{(c)} w^{(c)})_2^2 + \lambda |w| \tag{1}$$

Formulation (1) refers to the proposed formulation of centralized dsLassoCov for a regression problem. $\{x_i, y_i, x_i^{(c)}\}$ refer to the data pair of subject $i$, where $x_i \in R^p$ contains p features, $y \in R$ is the relevant outcome, and $x_i^{(c)} \in R^{1 \times p_c}$ contains $p_c$ covariates. $w \in R^{p \times 1}$ contains coefficients associating with features. And $w^{(c)} \in R^{p_c \times 1}$ contains the coefficients of $p_c$ covariates that need to be controlled for. Here, we assume that the effect of the covariates is linear. $x_i w$ is the linear prediction model. The equation $(y_i - x_i w - x_i^{(c)} w^{(c)})_2^2$ refers to the mean squared error (MSE) of subject i measuring the discrepancy between the predicted and actual outcome. $|w|$ is the sum of the absolute values of the coefficients and aims to exclude the features potentially associated with the covariates. $\lambda$ is the hyperparameter to control the number of selected features and can be estimated by a cross-validation procedure. In the paper, we only show the derivation of dsLassoCov regression algorithm as one can derive all equations for the classification algorithm in the same approach by switching to the logistic loss.

In the **supplementary methods**, we elucidate the technical details to solve the aforementioned objective, outline the training protocol and hyperparameters tuning, and detail the transformation of these methodologies into the federated learning algorithm.

*Covariate control mechanism*

To understand how dsLassoCov can control for associations of covariates, we explore the properties of the solution of dsLassoCov. Let $L(w^{(c)}{}_i)$ refered to the loss function presented in formulation (1). Due to its convexity, the first-order optimal condition, $\frac{dL(w^{(c)}{}_i)}{dw^{(c)}{}_i} = 0$ discribes the solution of $w^{(c)}$:

$$w^{(c)} = \left(x^{(c)T}x^{(c)}\right)^{-1} x^{(c)T}(y - xw)$$

Let $U_y^{(c)} = \left(x^{(c)T}x^{(c)}\right)^{-1} x^{(c)T}y$, $\hat{y} = xw$, and $U_{\hat{y}}^{(c)} = \left(x^{(c)T}x^{(c)}\right)^{-1} x^{(c)T}\hat{y}$. $U_y^{(c)}$ be interpreted as the coefficients regressed y on $x^{(c)}$ (e.g., $y = x^{(c)} \times U_y^{(c)}$), $\hat{y}$ is the estimator of y, and $U_{\hat{y}}^{(c)}$ contains the coefficients regressed $\hat{y}$ on $x^{(c)}$. Therefore,

$$w^{(c)} = U_y^{(c)} - U_{\hat{y}}^{(c)}$$

The objective function becomes

$$\min_w \frac{1}{2n} \sum_{i=1}^{n} || \left(y - x^{(c)}U_y^{(c)}\right) - \left(\hat{y} - x^{(c)}U_{\hat{y}}^{(c)}\right) ||_2^2 + \lambda|w| \quad (2)$$

Here $y - x^{(c)}U_y^{(c)}$ refers to the residual of y with the association of covariates $x^{(c)}$ removed. And $\hat{y} - x^{(c)}U_{\hat{y}}^{(c)}$ refers to the residual of $\hat{y}$ that with the association of covariates $x^{(c)}$ removed. The solution w is identified by minimizing the distance of the two residuals. This means the solution w is independent of associations with $x^{(c)}$. The formula (2) is analogous to a double machine learning approach[25], showcasing a debiased property. A similar approach has been applied to LASSO for controlling the effects of covariates[26].

*Simulation data analysis*

The simulation analysis aims to investigate the dsLassoCov algorithm's performance with regard to data dimensionality (efficiency), the number of servers (scalability), and the magnitude of covariates confounding effect (robustness). Key metrics used to assess the algorithm's performance include run-time and feature selection accuracy (performance of covariates control). Given that the main goal of our method is to identify potential risk factors in epidemiological studies, in our simulation analysis, we ignored the investigation of prediction performance. dsLassoCov was compared against two type of approaches: a federated baseline model and a meta-analysis of local ML approach. The federated baseline model (denoted as "ds.glm" (generalized linear model) for confounding effects removal followed by "ds.LASSO"), initially regresses out covariate effects from features and outcomes in a federated manner using multiple glm (one model per feature). Subsequently, a traditional federated LASSO model is trained on the corrected data for feature selection (all features into one single model). For the meta-analysis, four base machine learning models were employed: LASSO, ridge, support vector machine (SVM), and random forest. These models were trained locally on each server, and coefficients are aggregated to determine the selected features. Prior to model training, covariates are locally regressed on features and outcomes separately on each server to control confounding effects. To ensure the preservation of generalizability, both classification and regression tasks were simulated for all analyses. However, for classification tasks, covariates were only regressed out from features and not the outcome, due to the binary nature of the outcome. A detailed description of the machine learning workflow for the simulation data analysis is provided in the **supplementary methods**.

To support the aforementioned analysis, we simulate 20 features significantly associated with the outcome, among which 10 are confounded by covariates. The detailed data generation processes are outlined in the **supplementary methods**.

*Real data analysis*

This analysis is supported by the HELIX Project[18], which investigates early life environmental risk factors impacting lifelong health issues. As illustrated in **Figure 2**, data from the HELIX project are stored in six geographically distinct databases. A real DataSHIELD infrastructure, with each dataset hosted in a physically independent server, has been established among these databases to illustrate the usability of dsLassoCov in a real-world scenario.

The primary objective of this illustrative analysis was to identify early exposures associated with the diagnosis of hypertension. Early exposures in this context refer to environmental factors during pregnancy and the post-natal period, considered separately in the subsequent analyses. Two independent analyses were conducted, each involving 1143 healthy children and 155 children diagnosed with hypertension, distributed across the six data servers. In total, 77 exposures were included in the analysis, as detailed in **Table S2**. Additionally, 13 confounders which could potentially be associated with both the environmental exposures and blood pressure outcome were identified *a priori* based on previous literature[27,28], and considered in the analysis. Further information regarding the HELIX project, data cohorts, data preprocessing steps, and covariate inference can be found in the **supplementary methods**.

For each analysis (e.g., "pregnancy" or "post-natal"), we initially applied dsLassoCov under a 5-fold cross-validation procedure over a sequence of 100 lambda values to tune the hyperparameter. Subsequently, a final dsLassoCov model is trained on all data cohorts using the selected hyperparameter.

To demonstrate that the federated dsLassoCov model provides a comparable solution as observed in a non-federated environment, we compare the results of our federated dsLassoCov with the non-federated version of dsLassoCov, called "local-dsLassoCov", which was locally trained on the pooled dataset of all cohorts in a single computer. This local-dsLassoCov is comparable to running a Lasso model on all data together (pooled analysis) and was implemented based on the package glmnet[23].

*Code and data availability*

The client-side package (version of the package that needs tobe installed in the analyst computer): https://github.com/transbioZI/dsMTLClient

The server-side package (version of the package that needs to be : https://github.com/transbioZI/dsMTLBase

Tutorial: https://rpubs.com/aanguita/986397

**Results**

*Simulation data analysis*

In general, dsLassoCov consistently outperformed other approaches across all metrics for classification tasks. Regarding regression tasks, dsLassoCov demonstrated superior computational efficiency and indistinguishable feature selection ability compared to the conventional federated approach, (denoted as "glm (generalized linear model)" for confounding effects removal followed by "LASSO"), and outperformed meta-analysis approaches.

**Efficiency assessment**. **Figure 3** illustrates the results of the efficiency analysis comparing dsLassoCov against the conventional federated approach and meta-analysis approaches. For scenarios with low dimensionality (e.g., $\frac{\text{Features' number}}{\text{Subjects' number}} = 0.33$ ), the run-time of dsLassoCov is comparable to the conventional federated method "glm + LASSO". However, as data dimensionality increases, dsLassoCov exhibits gradually superior efficiency, as observed in **Figure 3** (a) and (c). Importantly, this superior performance is not attributed to a reduction in feature selection accuracy. **Figure 3** (b) and (d) highlight the increasing challenge of feature selection across all methods when the number of features in the dataset increases. Despite the challenge, dsLassoCov achieves higher (in classification settings) or indistinguishable (in regression settings) performance compared to other approaches. Specifically, in classification settings, dsLassoCov outperforms all other approaches, with some meta-analysis approach (e.g., SVM) following closely behind. In regression settings, as expected, dsLassoCov and "ds.glm + ds.lasso" perform comparably and better than meta-analysis approaches. Notably, compared to **Figure 3** (b), the results in (d) demonstrate that federated approaches in regression settings are more robust against higher dimensionality than in classification problems. For instance, for the regression task approximately 60% accuracy is still achieved when $\frac{\text{Features' number}}{\text{Subjects' number}}$ is equal to 10, while in classification it barely reaches a 30% of accuracy. This may be attributed to two factors: on the one hand, the continuous outcome in the regression task provides more information for model training, and on the other hand, it is the fact of having a binary outcome, which imply a limitation for all approaches relying on a first step of confounding removal through residualization (due to the fact that confounding effects can not be removed from a binary outcome through residualization), thereby inevitably confounding classification models.

**Scalability assessment. Figure 4** presents the results of the scalability analysis. In **Figure 4** (a) and (c), the run-time of dsLassoCov and the conventional approach "ds.glm + ds.Lasso" is depicted with an increasing number of servers. Notably, dsLassoCov outperformed "ds.glm + ds.Lasso" for any given number of servers. **Figure 4** (b) and (d) illustrate the comparison of feature selection accuracy on classification and regression tasks, respectively as the number of servers increases. Firstly, we can observe how all approaches achieve higher performance with an increasing number of servers. Secondly, in the classification setting, we can notice how dsLassoCov outperforms all other approaches for any number of servers. In the regression setting, dsLassoCov performed comparably to "ds.glm + ds.Lasso" and outperformed all meta-analysis approaches.

**Robustness assessment. Figure 5** presents the comparison of feature selection ability across different intensities of the confounding effect, denoted by α and γ, affecting outcomes and features, respectively. Consistent with previous findings, dsLassoCov outperforms "ds.glm + ds.Lasso" on classification tasks and exhibits comparable performance on regression tasks. Interestingly, it is observed that α has a stronger association with feature selection accuracy compared to γ. This observation can be attributed to the huge distorter effect that α elicits on associations in the simulated data. That is to say, our simulated outcomes comprises three effects: the true effect, the confounding effect, and an additional independent random effect. In scenarios with a high proportion of the confounding effect (high α value), the true effect is diminished, while the random effect remains unaffected. Consequently, the random effect becomes relatively amplified, posing a more challenging problem for feature selection.

*Real data analysis*

To demonstrate the applicability and usability of our dsLassoCov approach, we decided to conduct a real-world data analysis, involving a real dataset under a real DataSHIELD infrastructure. We leveraged the HELIX exposome dataset, which comprises one of the largest exposome resources for the pediatric population (~8 years old). Data from six independent cohorts (each dataset hosted in a different dataSHIELD server) were analyzed. To recap, our research question aimed to identify early-life environmental exposures associated with hypertension. The dsLassoCov model selected 32 exposures with an odd ratio distinct from (OR<0.95 or >1.05) for hypertension risk, as detailed in **Table S3** and illustrated in **Figure 6**. Among these selected exposures, the half were identified as risk factors for hypertension, encompassing outdoor exposures (such as building density), chemical exposures (such as PFAS), and lifestyle factors (such as exposure to a high-stress environment). The other half were highlighted as protective factors for hypertension and involved urban exposures such as the presence of public transport near to the area of residence. Notably, these findings replicate insights previously described[27,28], thereby reinforcing the reliability and trustability of dsLassoCov.

To investigate the comparability of our federated approach against the local traditional LASSO approach, we conducted LASSO analysis in a non-federated manner by merging data from all cohorts on the same computer and running the local-dsLassoCov. In this scenario, the reported optimal lambda value slightly differed (0.009) compared to federated dsLassoCov, resulting in a more sparse model. This discrepancy may be attributed to the optimization method employed by local traditional dsLassoCov, which favors selecting highly sparse solutions through coordinate descent[23]. However, this optimization approach is not feasible for federated learning adaptation due to its feature-wise nature, resulting in computationally intractable communication costs. Despite the different number of exposures selected, those chosen by local-dsLassoCov exhibited roughly the same estimated coefficients as those by federated dsLassoCov, as indicated in **Table S3**. Additionally, there were 13 exposures selected by federated dsLassoCov but ignored by local-dsLassoCov, with odds ratios close to 1. This minor discrepancy is deemed reasonable and may be attributed to sampling variation during the cross-validation procedure.

**Discussion**

In this manuscript, we introduce dsLassoCov, which, to the best of our knowledge, is the first federated learning algorithm designed to efficiently and effectively control confounding effects when conducting linear models in high-dimensional settings. dsLassoCov demonstrates strong robustness against statistical heterogeneity, a major challenge in federated learning research. Unlike other known countermeasures, such as personalization approaches[6] (which are designed for comprehensive federated learning applications with general heterogeneity), dsLassoCov focuses on addressing the widespread, long-standing, and complex heterogeneity (e.g., confounding effects) prevalent in the biomedical field. This heterogeneity poses a critical bottleneck as biomedical research transitions into a data-intensive and data-driven field. At present, our approach represents the only efficient and effective solution to this issue.

In the manuscript, we first provided a detailed exposition of the techniques underpinning dsLassoCov, including insights into its modeling, optimization, training protocol, hyperparameter tuning, and adaptation to the federated setting. Second, we offered a theoretical interpretation of the covariate control mechanism within dsLassoCov, illustrating how it implicitly executes covariate control procedure in each iteration. Third, despite the option to apply conventional FL methods for covariate control, our simulation data analysis revealed that dsLassoCov significantly outperforms others in terms of speed. Moreover, in classification tasks, it demonstrated superior performance, while in regression tasks, it at least matches the efficacy of other methods. Lastly, in our real data analysis, we showcase dsLassoCov's efficacy in capturing hypertension-associated early exposures, aligning well with previous research findings[27,28].

These results support the utility of dsLassoCov for biomedical applications, particularly in exposome studies where large datasets from multiple population cohorts are available. Recent trends in the field emphasize the importance of protecting sensitive patient data, as exemplified by initiatives like HELIX[18], ATHLETE[29] or LifeCycle projects[21], which serve as pioneering consortia for implementing federated analysis in epidemiological studies. In this context, dsLassoCov provides a federated learning solution for two major challenges encountered in exposome analysis: 1) high data dimensionality and 2) the presence of strong confounding effects and multicollinearity phenomena. Failing to address these challenges properly can lead to reduced power in detecting true associations and a higher likelihood of biased results, such as overestimating the effect of one exposure while underestimating another. dsLassoCov can address both situations and may therefore be a useful tool for effectively leveraging the resources generated by these and future consortia. By offering a robust solution to these challenges, dsLassoCov facilitates more accurate and reliable analysis of exposome data, ultimately advancing our understanding of environmental exposures and their impact on health outcomes.

Despite its utility, dsLassoCov does have limitations. Firstly, as demonstrated in simulation data analysis, dsLassoCov performed comparably to the "ds.glm + ds.Lasso" approach in terms of feature selection accuracy, indicating similar robustness against confounding effects. It's important to note that "ds.glm + dsLasso" is a fundamental and widely used approach in biomedical studies, e.g., molecular[30], exposome[27], clinical[31] studies, for covariate control. In this regard, dsLassoCov could potentially replace "ds.glm + dsLasso" in the era of federated learning due to its demonstrated superior efficiency in simulation analysis. Secondly, the current implementation only includes the linear LASSO model, limiting its application scenarios. However, it is theoretically feasible to simply extend the covariate-controlling mechanism to nonlinear prediction models by modifying the formulation (1). Nonetheless, due to the increased complexity of nonlinear models, additional theoretical development and validation work are required. Thirdly, dsLassoCov is integrated into the DataSHIELD[8] infrastructure, which currently supports only cross-silos federated learning applications, thereby limiting its usage in cross-device federated learning applications. Addressing this limitation would require further development in collaboration with DataSHIELD to expand into cross-device[6] federated learning applications.

In future work, while we have demonstrated the adaptability of complex models within the dsLassoCov framework without compromising the covariates removal mechanism, a quantitative investigation is warranted to determine the performance of complex dsLassoCov in terms of model complexity, runtime, and feature selection accuracy.

**Conclusion**

We introduce dsLassoCov, a pioneering federated learning method designed to perform LASSO regression while addressing confounding effects. Positioned as a novel approach in the realm of federated learning, dsLassoCov stands out as a specialized solution tailored to combat a significant source of heterogeneity: confounding effects prevalent in various biomedical applications. With its unique focus on confounding effects, dsLassoCov has the potential to serve as a foundational element in transitioning numerous biomedical studies into the realm of large-scale data-driven analyses.

**Acknowledgements**

This publication was supported through state funds approved by the State Parliament of Baden-Württemberg for the Innovation Campus Health + Life Science alliance Heidelberg Mannheim. ATHLETE project has received funding from the European Union's Horizon 2020 research and innovation program under grant agreement no 874583. This publication reflects only the authors' view, and the European Commission is not responsible for any use that may be made of the information it contains. We acknowledge support from the grant CEX2018-000806-S funded by MCIN/AEI/10.13039/501100011033 and support from the Generalitat de Catalunya through the CERCA Program. We also thank support from the grant FJC2021-046952-I funded by MCIN/AEI/10.13039/501100011033 and by "European Union NextGenerationEU/PRTR" and acknowledge funding from the Ministry of Research and Universities of the Government of Catalonia (2021-SGR-01563).

# References

1. Sun, S., Wang, C., Ding, H. & Zou, Q. Machine learning and its applications in plant molecular studies. *Briefings in functional genomics* **19**, 40-48 (2020).
2. Watson, D.S. *et al.* Clinical applications of machine learning algorithms: beyond the black box. *Bmj* **364**(2019).
3. Asiimwe, R. *et al.* From biobank and data silos into a data commons: convergence to support translational medicine. *Journal of Translational Medicine* **19**, 1-13 (2021).
4. Voigt, P. & Von dem Bussche, A. The eu general data protection regulation (gdpr). *A Practical Guide, 1st Ed., Cham: Springer International Publishing* **10**, 10-5555 (2017).
5. Decherchi, S., Pedrini, E., Mordenti, M., Cavalli, A. & Sangiorgi, L. Opportunities and challenges for machine learning in rare diseases. *Frontiers in medicine* **8**, 747612 (2021).
6. Mammen, P.M. Federated learning: Opportunities and challenges. *arXiv preprint arXiv:2101.05428* (2021).
7. Doiron, D. *et al.* Data harmonization and federated analysis of population-based studies: the BioSHaRE project. *Emerg Themes Epidemiol* **10**, 12 (2013).
8. Wilson, R.C. *et al.* DataSHIELD – New Directions and Dimensions. *Data Science Journal* **16**(2017).
9. Nasirigerdeh, R. *et al.* sPLINK: A Federated, Privacy-Preserving Tool as a Robust Alternative to Meta-Analysis in Genome-Wide Association Studies. *BioRxiv* (2020).
10. Marcon, Y. *et al.* Orchestrating privacy-protected big data analyses of data from different resources with R and DataSHIELD. *PLoS Comput Biol* **17**, e1008880 (2021).
11. Wang, J. & Ma, F. Federated learning for rare disease detection: a survey. *Rare Disease and Orphan Drugs Journal* **16**(2023).
12. Li, T. *et al.* Federated optimization in heterogeneous networks. *Proceedings of Machine learning and systems* **2**, 429-450 (2020).
13. Leek, J.T. *et al.* Tackling the widespread and critical impact of batch effects in high-throughput data. *Nature Reviews Genetics* **11**, 733-739 (2010).
14. Jager, K.J., Zoccali, C., Macleod, A. & Dekker, F.W. Confounding: what it is and how to deal with it. *Kidney Int* **73**, 256-60 (2008).
15. Snoek, L., Miletić, S. & Scholte, H.S. How to control for confounds in decoding analyses of neuroimaging data. *Neuroimage* **184**, 741-760 (2019).
16. Brenner, H. & Blettner, M. Controlling for continuous confounders in epidemiologic research. *Epidemiology* **8**, 429-434 (1997).
17. Frisch, R. & Waugh, F.V. Partial time regressions as compared with individual trends. *Econometrica: Journal of the Econometric Society*, 387-401 (1933).
18. Maitre, L. *et al.* Human Early Life Exposome (HELIX) study: a European population-based exposome cohort. *BMJ Open* **8**, e021311 (2018).
19. Cao, H. *et al.* dsMTL - a computational framework for privacy-preserving, distributed multi-task machine learning. *Bioinformatics* (2022).
20. Gaye, A. *et al.* DataSHIELD: taking the analysis to the data, not the data to the analysis. *International journal of epidemiology* **43**, 1929-1944 (2014).
21. Jaddoe, V.W.V. *et al.* The LifeCycle Project-EU Child Cohort Network: a federated analysis infrastructure and harmonized data of more than 250,000 children and parents. *Eur J Epidemiol* **35**, 709-724 (2020).
22. Tibshirani, R. Regression shrinkage and selection via the lasso. *Journal of the Royal Statistical Society. Series B (Methodological)*, 267-288 (1996).
23. Zou, H. & Hastie, T. Regularization and variable selection via the elastic net. *Journal of the Royal Statistical Society: Series B (Statistical Methodology)* **67**, 301-320 (2005).

24. Tibshirani, R. Regression shrinkage and selection via the lasso: a retrospective. *Journal of the Royal Statistical Society: Series B (Statistical Methodology)* **73**, 273-282 (2011).
25. Chernozhukov, V. *et al.* Double/debiased machine learning for treatment and structural parameters. (Oxford University Press Oxford, UK, 2018).
26. Urminsky, O., Hansen, C. & Chernozhukov, V. Using Double-Lasso Regression for Principled Variable Selection. *SSRN Electronic Journal* (2016).
27. Warembourg, C. *et al.* Urban environment during early-life and blood pressure in young children. *Environ Int* **146**, 106174 (2021).
28. Warembourg, C. *et al.* Early-Life Environmental Exposures and Blood Pressure in Children. *J Am Coll Cardiol* **74**, 1317-1328 (2019).
29. Vrijheid, M. *et al.* Advancing tools for human early lifecourse exposome research and translation (ATHLETE): Project overview. *Environ Epidemiol* **5**, e166 (2021).
30. Cao, H., Chen, J., Meyer-Lindenberg, A. & Schwarz, E. A polygenic score for schizophrenia predicts glycemic control. *Transl Psychiatry* **7**, 1295 (2017).
31. Elze, M.C. *et al.* Comparison of propensity score methods and covariate adjustment: evaluation in 4 cardiovascular studies. *Journal of the American College of Cardiology* **69**, 345-357 (2017).

**Figures**

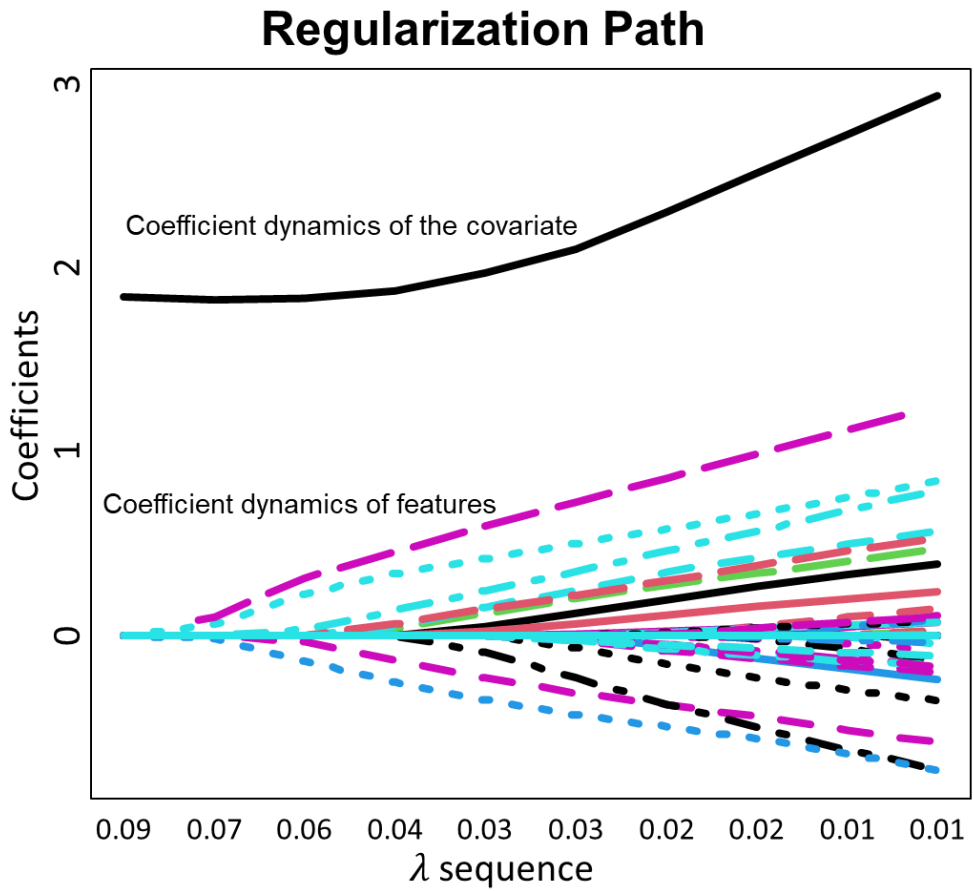

**Figure 1. The regularization path of dsLassoCov.** The regularization path refered to outputs of dsLassoCov model training, creating a series models indexed by a lambda sequence. By reducing the lambda value, a larger model with more features is obtained. The coefficient of covariate is freely varied along the sequence and aimed to suppress the selection of covariate-associated features.

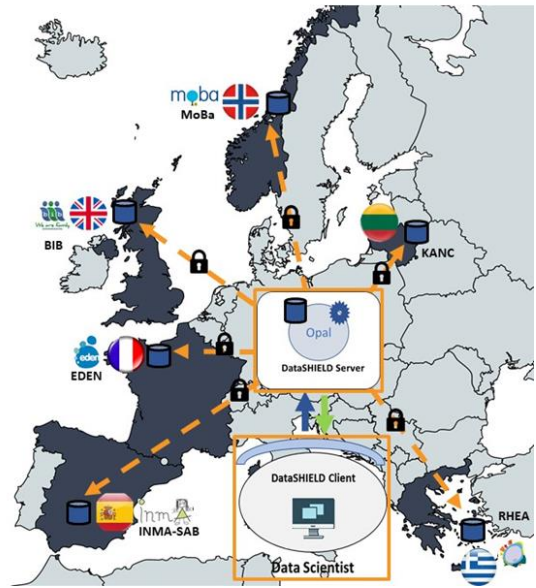
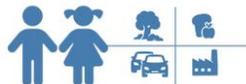
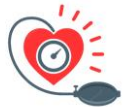

**Figure 2. Input HELIX datasets organized by cohort in the Opal server.**

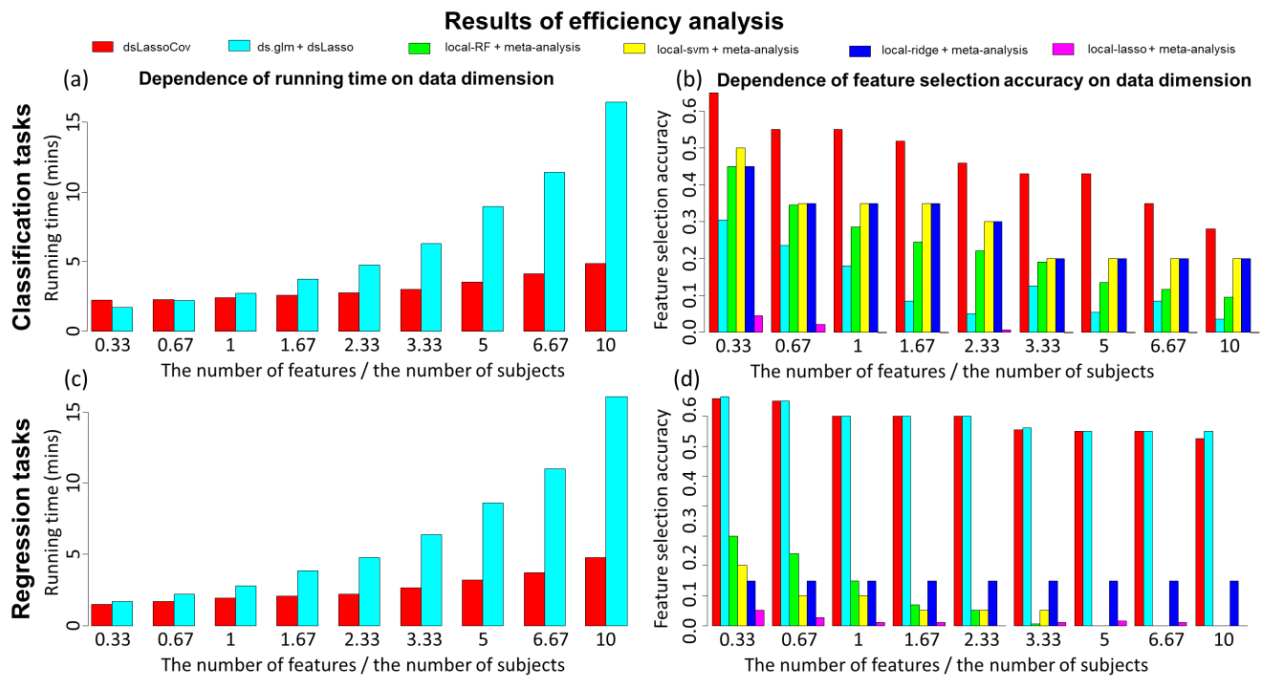

**Figure 3. Result of the efficiency analysis.** The dependence of the algorithm's performance on the data dimension is explored for both regression and classification tasks. The run time and feature selection accuracy are used as metrics. Three servers were simulated in this analysis, and each server contained

100 subjects. The dimensionality increased from 100 to 3000, therefore the ratio $\frac{Features'number}{Subjects'number}$ ranged from 0.33 to 10. Panel (a)~(b) shows the results of classification tasks. Panel (c)~(d) shows the results of regression tasks. Panel (a) and (c) showed dsLassoCov is more efficient than the conventional federated approach in particular for high-dimensional problem. As shown in (b) and (d), meta-analysis of different local ML models achieved suboptimal feature selection accuracy. And dsLassoCov outperformed other approaches in classification task, and is comparable to conventional federated approach in regression setting.

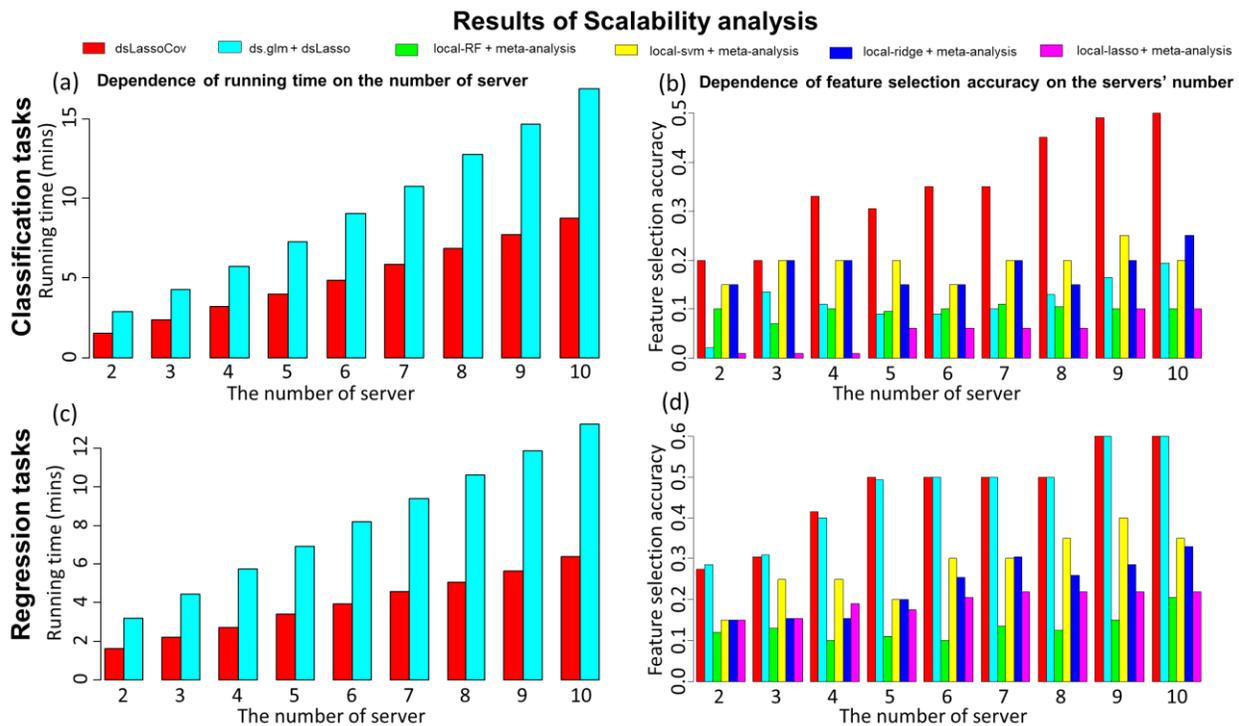

**Figure 4. Result of the scalability analysis.** The dependence of the algorithm's performance on the number of servers was explored in classification and regression setting. The run time and feature selection accuracy are used as metrics. Up to ten servers were simulated in this analysis, and each server contained 50 'subjects' with 1000 features. An increasing number of servers is gradually used in the analysis, and the run time as well as feature selection accuracy, are recorded for comparison. Panel (a) shows the comparison of the run time among federated approaches. Panel (a)~(b) shows the results of classification tasks. Panel (c)~(d) shows the results of regression tasks. Panel (a) and (c) illustrated dsLassoCov is more efficient than the conventional federated approach. Panel (b) showed dsLassoCov a superior feature selection accuracy among all methods in classification setting. Panel (d) showed a comparative performance of dsLassoCov and conventional federated approach in regression setting.

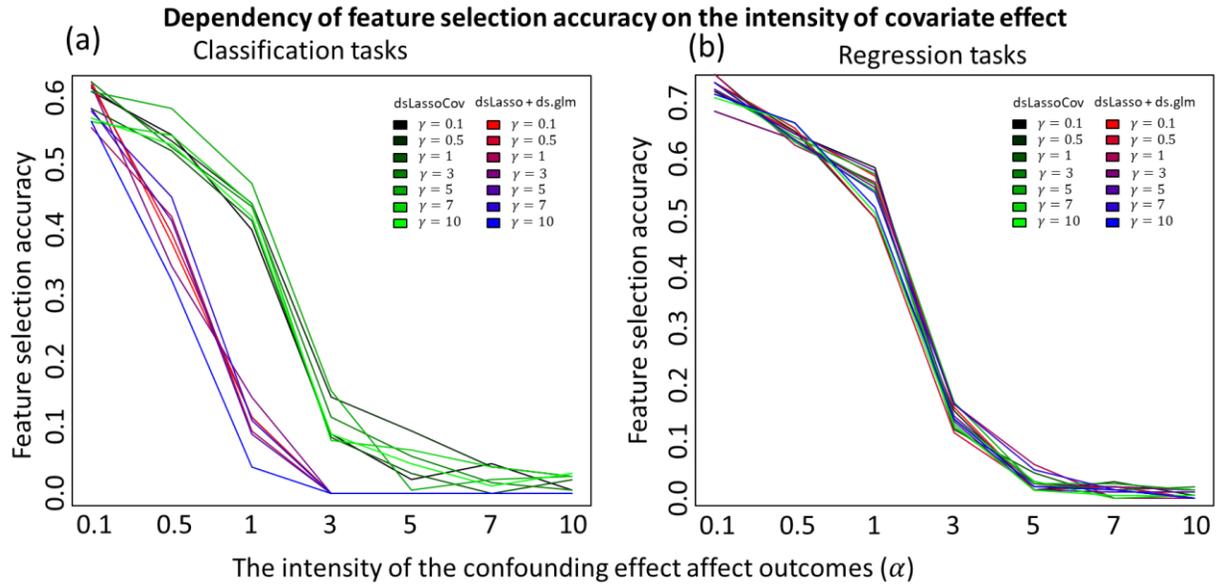

**Figure 5, Result of the robustness analysis.** The dependence of the feature selection accuracy on the intensity of confounding effects was explored in classification and regression setting. $\alpha$ and $\gamma$ referred to the intensities of the confounding effects affecting on outcomes and features, respectively. As the comparison, the conventional federated approach has been applied in the same setting. Three servers are simulated, and each contained 100 subjects with 1000 features. The parameter pair $\{\alpha, \gamma\}$ is varied from 0.1 to 10, and the simulated data is created for each variation. Panel (a) and (b) shows the results of classification and regression tasks. dsLassoCov outperformed the conventional federated approach in the classification setting, and performed indiscriminately in the regression setting. Moreover, compared to $\gamma$, $\alpha$ determines more the influence to feature selection accuracy.

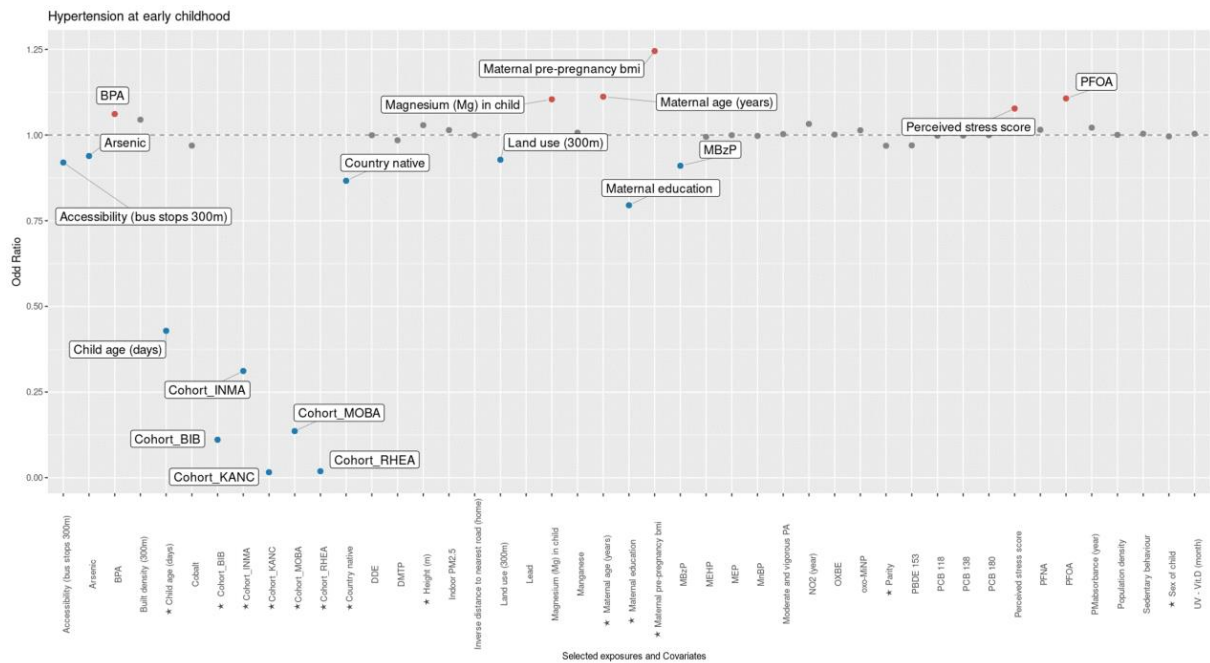

**Figure 6, Estimated odd ratios for selected exposures in the real HELIX dataset (Model: HT ~ exposome + confounders).** Only exposures showing an odd ratio higher than 1.05 or lower than 0.95 are plotted. Confounders are identified here with an astherics (*).

# Supplementary Methods of "dsLassoCov: a federated learning approach incorporating covariate control"

Outlines:

- Modelling, optimization, training protocol and algorithm
    - Modeling and optimization
    - non-federated training protocol
    - Federated training protocol
    - Federated learning algorithm
- Simulation data analysis
    - The competing methods
    - Machine learnming workflow
    - Confounded data simulation
    - Efficiency assessment
    - Scalability assessment
    - Robustness assessment
- Real data analysis
    - Helix project
    - Data cohorts and preprocessing
    - Covariate inference
- Introduction of DataSHIELD

# Modelling, optimization and algorithm

*The modeling and optimization*

Let's consider the linear prediction model f(x, w) = xw, the loss function becomes,

$$\min_{w,w_c} \frac{1}{2n} \sum_{i=1}^{n} (y_i - x_i w - x_i^{(c)} w^{(c)})_2^2 + \lambda |w| \tag{1}$$

Formulation (1) refers to the loss function of centralized dsLassoCov for a regression problem. $\{x_i, y_i, x_i^{(c)}\}$ refered to the data pair of subject $i$, where $x_i \in R^p$ contains p features, $y \in R$ is the relevant outcome, and $x_i^{(c)} \in R^{1 \times p_c}$ contains $p_c$ covariates. $w \in R^{p \times 1}$ contains coefficients associating with features. And $w^{(c)} \in R^{p_c \times 1}$ contains the $p_c$ covariates that need to be controlled for. The equation $(y_i - x_i w - x_i^{(c)} w^{(c)})_2^2$ refers to the least-squared loss of subject i measuring the discrepancy between the predicted and actual outcome. $|w|$ is the "Lasso-like" regularization term[22] and aims to exclude the features potentially associated with the covariates. $\lambda$ is the hyperparameter to control the number of selected features and can be estimated by a cross-validation procedure. In the paper, we only show the derivation of dsLassoCov regression algorithm as one can derive all equations for the classification algorithm in the same approach by switching to the logistic loss.

Let $F(w, w^{(c)}) = \frac{1}{2n} \sum_{i=1}^{n} (y_i - x_i w - x_i^{(c)} w^{(c)})_2^2$ and $\Omega(w) = |w|$, the formulation (1) can be re-written as problem (2),

$$\min_{w,w^{(c)}} F(w, w^{(c)}) + \lambda \Omega(w) \tag{2}$$

$F(w, w^{(c)})$ is a smooth and convex function of $w$ and $w^{(c)}$, and $\Omega(w)$ is non-smooth and convex. The penalty is only performed on w. Assume the Lipschitz constant of $F(w, w^{(c)})$ is $L$, a critical step is to solve the following subproblem (3) in each iteration,

$$w_{i+1} = \arg\min_{y} F(w_i, w^{(c)}_i) + \langle \nabla_{w_i} F(w_i, w^{(c)}_i), y - w_i \rangle + \frac{L}{2} \|y - w_i\|_2^2 + \lambda \Omega(y)$$

and

$$w^{(c)}_{i+1} = \arg\min_{y} F(w_i, w^{(c)}_i) + \langle \nabla_{w^{(c)}_i} F(w_i, w^{(c)}_i), y - w^{(c)}_i \rangle + \frac{L}{2} \|y - w^{(c)}_i\|_2^2 \tag{3}$$

Problem (3) is the second-order approximation of $F(.)$ at the current estimators $w_i$ and $w^{(c)}_i$. $w_{i+1}$ and $w^{(c)}_{i+1}$ refers to the updated estimates of coefficients of features and covariates, respectively. After re-organization, we have an equivalent form (4),

$$w_{i+1} = \arg\min_{y} \frac{L}{2} \left( y - \left( w_i - \frac{\nabla_{w_i} F(w_i, w^{(c)}_i)}{L} \right) \right)^2 + \lambda |y|$$

and

$$w^{(c)}_{i+1} = \arg\min_{y} \frac{L}{2} \left( y - \left( w^{(c)}_i - \frac{\nabla_{w^{(c)}_i} F(w_i, w^{(c)}_i)}{L} \right) \right)^2 \tag{4}$$

where the gradients of $F(.)$ are

$$\nabla_{w^{(c)}_i} F(w_i, w^{(c)}_i) = \frac{1}{n} \sum_{m=1}^{n} x_m^{(c)T}(x_m w_i + x_m^{(c)} w^{(c)}_i - y_m)$$

and (5)

$$\nabla_{w_i} F(w_i, w^{(c)}_i) = \frac{1}{n} \sum_{m=1}^{n} x_m^{T}(x_m w_i + x_m^{(c)} w^{(c)}_i - y_m)$$

Investigating the equations (4), $w^{(c)}_{i+1}$ can be solved analytically, as shown in (6).

$$w^{(c)}_{i+1} = w^{(c)}_i - \frac{\nabla_{w^{(c)}_i} F(w_i, w^{(c)}_i)}{L} \quad (6)$$

For $w_{i+1}$ in problem (4), the expression is separable and can be divided into $p$ independent small problems to solve. Let $w^{(j)}_i \in \mathbb{R}$ is the $j$th value of the vector $w_i$, each small problem takes the form (7),

$$w^{(j)}_{i+1} = \arg\min_{y^{(j)}} \frac{L}{2}\left(y^{(j)} - \tilde{w}^{(j)}_i\right)^2 + \lambda |y^{(j)}|$$

where

$$\tilde{w}^{(j)}_i = w^{(j)}_i - \frac{\nabla_{w_i}(w_i, w^{(c)}_i)^{(j)}}{L} \quad (7)$$

This problem (7) can be solved analytically

$$w^{(j)}_{i+1} = \max\left\{\tilde{w}^{(j)}_i - \frac{\lambda}{L}, 0\right\} \times \text{sign}(\tilde{w}^{(j)}_i) \quad (8)$$

Since the Lipschitz constant L is usually unknown, we estimated it using the Ajimo-Goldstein line search. Since the objective of dsLassoCov is convex and non-smooth, we accelerate the entire procedure with Nesterov's acceleration approach [32-34]. For each iteration $i$, a search point is quantified by weighing the estimates from the previous two iterations,

$$s_i = \frac{\alpha_{i-1}}{\alpha_i} w_i + \frac{1 - \alpha_{i-1}}{\alpha_i} w_{i-1} \quad (9)$$

The search point is subsequently sent to (6) ~(8) to update the solution. The procedure will be repeated until the solution converges to the required precision.

*The non-federated training protocol*

Instead of estimating a model for each specific λ (lambda), we derived the entire regularization path across a sequence of λ values, organized in descending order. This sequence represents a range of potential hyperparameter values, where a higher λ is associated with fewer identified features. Such an approach enhances statistical power by focusing the analysis on a select number of features that are most relevant to the outcomes, thereby emabling the covariate control procedure working.

The training protocol including the estimation of the λ sequence, has previously been explored[19,23]. A critical aspect of this approach is determining an optimal λ sequence that is sufficiently comprehensive to capture the highest likelihood of relevant features while minimizing the inclusion of superfluous ones. To establish this optimal sequence, we utilized a strategy inspired by the glmnet[23] procedure, beginning with the calculation of the largest λ (λ_max) as the starting point of the regularization path. The smallest λ in the sequence is then set relative to λ_max (e.g., λ_max*0.01), with the complete sequence being interpolated on a logarithmic scale.

In the context of the dsLassoCov model, determining λ_max requires satisfying specific equations that stem from the first-order optimal conditions of our **formulation (1).**

$$\nabla_w F(w, w^{(c)}) + \lambda_{p\times 1} \sigma_{p\times 1} = 0, \sigma \in [-1, 1]$$
$$\nabla_{w^{(c)}} F(w, w^{(c)}) = 0 \qquad (10)$$
$$w = 0$$

After solving equations in (10), we have

$$\lambda_{p\times 1} - \frac{1}{n} \left| \frac{x^T y - x^T x^{(c)} \left( x^{(c)^T} x^{(c)} \right)^{-1} x^{(c)^T} y}{\sigma_{p\times 1}} \right| = 0$$

Therefore, the smallest possible value of lambda is linked to $\sigma = 1$ or $-1$. Therefore,

$$\lambda_{\max} = \max \left( \frac{1}{n} \left| x^T y - x^T x^{(c)} \left( x^{(c)^T} x^{(c)} \right)^{-1} x^{(c)^T} y \right| \right)_i, i \in [1, p]$$

With λ_max determined, we can interpolate the entire λ sequence as previously described, enabling the training of a series of sparse models utilizing the "warm-start" technique detailed in our prior work[19]. In a federated setting, where data reside across multiple remote servers, adapting these calculations to a distributed environment is crucial. Our adaptation strategy ensures that the federated model's solution is exactly congruent with that of its non-federated counterpart.

*The federated training protocol*

Our federated algorithm leverages the DataSHIELD[20] infrastructure, wherein data are securely stored on DataSHIELD servers and interact exclusively with the DataSHIELD Client. Specifically, for the dsLassoCov algorithm, two essential functions are deployed on the server-side: one to access the gradient of the loss function F(·), using estimated parameters and local data, and another to evaluate F(·) itself. These functions are invoked during runtime, with their results transmitted back to the client. On the client side, values from all servers are aggregated to ascertain the final gradient and function value, as outlined below.

Crucially, to ensure the identification of an unbiased solution, we opt for calculating the exact gradients and function values rather than resorting to statistical approximations. While approximations could potentially enhance the efficiency of federated learning methods[35], precision in these calculations is paramount for our purposes. The specific calculations employed are detailed subsequently.

$$\text{dist } \nabla_w F(w, w^{(c)}) = \frac{1}{n} \sum_{i=1}^{t} \nabla_w F(w, w^{(c)}; D^{(i)}) \times n_t$$

$$\text{dist } \nabla_{w^{(c)}} F(w, w^{(c)}) = \frac{1}{n} \sum_{i=1}^{t} \nabla_{w^{(c)}} F(w, w^{(c)}; D^{(i)}) \times n_t$$

$$\text{dist } O(w_{i+1}, w^{(c)}{}_{i+1}) = \frac{1}{n} \sum_{i=1}^{t} F(w, w^{(c)}; D^{(i)}) \times n_t$$

*Federated learning algorithm*

**Algorithm 1** summarizes the algorithm performed on the client side to solve the objective, wherein specific functions starting with "dist" required the remote access to the servers and aggregated the resturns. The estimation of the regularization path is summarized in Algorithm 2. Prior to Algorithm 2, the lambda sequence is estimated as shown above.

Based on these functions, we also implemented the cross-validation procedure in dsLassoCov to select the optimal lambda. All these functionalities for the classification algorithm are also provided.

**Algorithm 1** encapsulates the client-side operations required to solve our optimization problem, notably incorporating functions prefixed with "dist" to enable the remote execution of data-relevant operations on servers access and results aggregation. The detailed steps for models' estimation are encapsulated in **Algorithm 2**, which follows the prior determination of the lambda sequence as discussed. In the practical application, to determine the lam_max, we applied ds.glm() function[20] in the DataSHIELD.

Leveraging this and other specific functions, we have integrated a cross-validation procedure within the dsLassoCov package to pinpoint the optimal lambda value. All these functionalities for the classification algorithm are also provided.

---

**Algorithm 1 dsLassoCov client-side algorithm**

**Input**: $\lambda > 0, L_0 > 0, w_0, w^{(c)}_0$ $maxIter > 0$

**Output**: $w_{i+1}, w^{(c)}_{i+1}$

1: Initialize $w_1 = w_0, w^{(c)}_1 = w^{(c)}_0, \alpha_{-1} = \alpha_0 = 0$, and $L = L_0$

2: **for** $i = 1$ to $maxIter$ **do**

3:      Set $S_i = w_i + \frac{\alpha_{i-1}-1}{\alpha_i}(w_i - w_{i-1})$

4:      Set $S^{(c)}_i = w^{(c)}_i + \frac{\alpha_{i-1}-1}{\alpha_i}(w^{(c)}_i - w^{(c)}_{i-1})$

5:      **for** $L$ in $\{L_{i-1}, 2L_{i-1}, 4L_{i-1}, 16L_{i-1}, \ldots\}$ **do**

6:          $G_w = \text{dist } \nabla_w F(S_i, S^{(c)}_i)$

7:          $G_{w^{(c)}} = \text{dist } \nabla_{w^{(c)}} F(S_i, S^{(c)}_i)$

8:          $w_{i+1} = \max\left\{S_i - \frac{G_w - \lambda}{L}, 0\right\} \times \text{sign}(S_i - \frac{G_w}{L})$

9:          $w^{(c)}_{i+1} = S^{(c)}_i - \frac{\text{dist } \nabla_{w^{(c)}} F(S_i, S^{(c)}_i)}{L}$

10:         $\mathcal{M} = F(w_i, w^{(c)}_i) + \langle G_w, w_{i+1} - w_i \rangle + \langle G_{w^{(c)}}, w^{(c)}_{i+1} - w^{(c)}_i \rangle + \frac{L}{2}\left\|w^{(c)}_{i+1} - w^{(c)}_i\right\|^2_2 + \frac{L}{2}\left\|w_{i+1} - w_i\right\|^2_2$

11:         **if** $\mathcal{M} \geq \text{dist } O(w_{i+1}, w^{(c)}_{i+1})$ **do**

                 **Break**

            **end if**

     **end for**

12:      Set $L_i = L$, and $\alpha_{i+1} = \frac{1 + \sqrt{1 + 4\alpha_i^2}}{2}$

13:      If termination rule satisfied, **return**

   **end for**

---

**Algorithm 2** Estimation procedure of regularization path

**Input**: $\lambda_1 > \lambda_2 > \cdots > 0$

**Output**: $\{w_1, w^{(c)}_1\}, \{w_2, w^{(c)}_2\}, \{w_3, w^{(c)}_3\}, \ldots$

1: Initialize $w_0 = p \times 1 = 0, w^{(c)}_0 = p^{(c)} \times 1 = 0$,

2: **for** $i = \{1, 2, \ldots\}$ **do**

3:      $W_i =$**Algorithm 1** ($\lambda = \lambda_i, L_0 = 1, w_0 = w_{i-1}, w^{(c)}_0 = w^{(c)}_{i-1}, maxIter = 100$)

## Simulation data analysis

We conducted a simulation-based data analysis to demonstrate the efficacy, scalability, and robustness of our novel method. Efficiency analysis is focused on quantifying the algorithm's performance as the data dimensionality increases. Scalability analysis, on the other hand, aims to evaluate how the algorithm performs as the number of servers utilized grows. Lastly, robustness analysis delves into understanding how varying degrees of confounding effects impact the algorithm's effectiveness. The primary metrics employed for evaluating the algorithm's utility include runtime and feature selection accuracy.

*The competing methods*

To facilitate a comprehensive comparison of runtime, we employ the conventional federated approach, termed "ds.glm + dsLasso", as our baseline. This approach involves two main steps: first, utilizing ds.glm() from the dsBaseClient[10] package to execute a federated generalized linear model, which aims to mitigate covariate effects from the features and outcomes (known as "regressing out covariates"[15] approach). Subsequently, dsLasso() from the dsMTL[19] package is employed to train a federated Lasso model. The simulation of servers is executed on a single machine using the dsLite[10] package.

For assessing feature selection accuracy, the "ds.glm + dsLasso" and meta-analysis are used as a comparison. In the meta-analysis, four distinct machine learning models—namely, Random Forest (RF), Support Vector Machine (SVM), Lasso, and Ridge Regression—are trained locally on each server. These models are then aggregated to identify the crucial features. This comparison enables a comprehensive evaluation of feature selection accuracy across different methodologies.

*Machine learnming workflow*

We perform a 5-fold cross-validation first to select the hyperparameters (if there are any) and then a model training to obtain the coefficients. For feature selection, we rank the non-zero absolute coefficients and take up to 20 from the top as the selected features. The successful selection rate is returned as the selection accuracy. For the meta-analysis, we investigated the performance of common machine learning methods, logistic regression[23], ridge regression[23], randomforest[36] and svm[37]. For ridge, logistic regression and svm models, the coefficients returned from servers are aggregated by the average of absolute values. The results are ranked, and up to 20 from the top are non-zero are selected. For the random forest model, instead of using coefficients, we rank the "importance" of features for selection.

*Confounded data simulation*

We considered that the confounding effect affected the data generation procedure of the outcomes and the confounded features. This will lead to the identification of spuriously outcome-associated (non-significant) features if the covariates are not controlled for. We constructed the simulation data to quantify how well dsLassoCov could capture the true signals by controlling the covariates.

The inputs are the number of subjects ($n$), the number of features ($p$), the number of confounded features ($p_c$), the number of true features ($p_t$), the intensity of the confounder affecting the outcome ($a$) and the intensity of the confounder affecting the features ($\gamma$).

First, the confounding effect $c \in R^{n \times 1}$ is simulated, each term is created as $\text{sign}(\mathcal{N}(0,1)) \times \mathcal{N}(1,1)$. Second, the coefficient vector of true features $w \in R^{p_t \times 1}$ is created, $w_i = \text{sign}(\mathcal{N}(0,1)) \times \mathcal{N}(1,1)$. Third, the design matrix $X \in R^{n \times p}$ is created as $X = \mathcal{N}(0,1)^{n \times p} = [X_{sig} \quad X_c \quad X_{non}]$, where $X_{sig} \in R^{n \times p_t}$ contained the true features, $X_c \in R^{n \times p_c}$ contained the non-significant features that will be confounded, and $X_{non} \in R^{n \times (p - p_c - p_t)}$ contained the outcome-irrelevant features.

Then the process that the non-significant features are affected by the confounding effects is simulated as,

$$X_c^{(j)} \leftarrow X_c^{(j)} + \gamma c \tag{11}$$

$X_c^{(j)}$ referred to the jth column of $X_c$. Then the z-standardization is applied to each feature of $X_c$ and $X_{sig}$.

The generation of outcomes is followed.

$$Y = [X_{sig} \quad X_c \quad a] \begin{bmatrix} w \\ 0 \\ c \end{bmatrix} \tag{12}$$

Then the z-standardization is applied to $Y$. Here, $a$ and $\gamma$ are scalars representing the intensities of confounding effect affecting on the outcome and features, respectively.

For the regression setting, $Y \leftarrow Y + \epsilon$, where $\epsilon = \mathcal{N}(0, 0.5)^{n \times 1}$. For classification setting, $Y \leftarrow \text{sign}(Y + \epsilon)$, where $\epsilon = \mathcal{N}(0, 0.1)^{n \times 1}$.

*Efficiency assessment*

In our efficiency test, our objective was to measure how the runtime of the algorithm correlates with the increasing dimensionality of the data. The simulation process adhered to **equations (11)** and **(12)** as outlined in our methodology. We simulated three servers, each housing the feature data of 100 'subjects' along with their corresponding outcomes, which could be binary or continuous. Among these features, 20 were genuinely associated with the outcomes. Additionally, we introduced a single confounding effect that influenced 10 spuriously outcome-associated features. The intensity parameters governing the confounding effects on both outcomes and features were set to 1. We systematically varied the feature dimension, selecting values sequentially from the set $\{100, 200, 300, 500, 700, 1000, 1500, 2000, 3000\}$. The parameters γ and α were fixed at 1 for this experiment.

*Scalability assessment*

In our scalability test, our objective was to assess how the runtime of the algorithm varies with an increasing number of servers. Following the procedure outlined in **equations (11)** and **(12)**, we conducted our simulations. A total of 10 servers were simulated for this experiment. Each server housed the feature data of 50 independent 'subjects' along with their corresponding outcomes, which could be binary or continuous. The feature dimension was fixed at 1000, with 20 features truly significant to the outcomes.

Additionally, we introduced a single confounding effect that influenced 10 spuriously outcome-associated features. The parameters γ and α were set to 1 for this experiment. This setup allowed us to systematically explore the impact of scaling the number of servers on algorithm runtime.

*Robustness assessment*

In our robustness test, our goal was to assess how the intensity of the confounding effect impacts the algorithm's feature selection capability. To conduct this analysis, we generated feature data and outcomes (which could be binary or continuous) for 300 'subjects'. These subjects were evenly distributed across three simulated servers, with each subject possessing 1000 features.

We systematically varied the parameters $a$ and $\gamma$, where $a \in \{0.1, 0.5, 1, 3, 5, 7, 10\}$ and $\gamma \in \{0.1, 0.5, 1, 3, 5, 7, 10\}$, representing the intensity of the confounding effects affecting on the coutcome and features. For every combination of these two parameters, we trained both the dsLassoCov and "ds.glm + dsLasso" algorithms. Subsequently, we compared the resulting feature selection accuracy. This comprehensive examination allowed us to evaluate the algorithm's robustness across various levels of confounding effects.

# Read data analysis

To showcase the practical utility of dsLassoCov in real-world multi-center studies, we opted for complex exposome data. Our focus was on utilizing dsLassoCov to identify early environmental exposures that potentially influence the risk of hypertension.

*Helix project*

The data cohorts investigated here are derived from the HELIX (Human Early-Life Exposome) Project[18]. The HELIX project gathers data from 6 longitudinal-based European birth cohorts with the aim of evaluating the effect of environmental risk factors on the health of mothers and children. We used data from the Helix sub-cohort consisting of data from approximately 200 children in each cohort (Table S1). In the 1,298 children, a wide range of environmental exposures were evaluated to define the early-life exposome during two time periods: the prenatal pregnancy period and the postnatal period (childhood age 6 to 11 years). Exposures were estimated based on the home and school address (provided details here on in the supplement). A general description of the study population can be found in the Supplementary **Table S1**, and a summary of the structure and organization of the dataset in the federated scenario is illustrated in **Figure 2**.

The data cohorts under investigation stem from the HELIX (Human Early-Life Exposome) Project. This initiative aggregates data from six longitudinal-based European birth cohorts with the overarching objective of assessing the impact of environmental risk factors on the health outcomes of both mothers and children. Specifically, we utilized data from the Helix sub-cohort, which comprises approximately 200 children from each cohort (refer to **Table S1** for details). Within a cohort of 1,298 children, an extensive array of environmental exposures was scrutinized to delineate the early-life exposome across two distinct time periods: the prenatal pregnancy period and the postnatal period spanning childhood ages 6 to 11 years. These exposures were assessed based on information obtained from both home and school addresses (further elaborated in the supplement). For a comprehensive understanding of the study population, please refer to **Table S1**. Furthermore, **Figure 2** provides a succinct depiction of the structure and organization of the dataset within the federated scenario.

*Data cohorts and preprocessing*

Environmental exposures used in this analysis consisted of outdoor exposures, chemical exposures and lifestyle factors. Outdoor exposures included built environment, meteorological conditions, natural spaces, road traffic, and noise. Chemical exposures included: organochlorine compounds (including polychlorinated biphenyls [PCBs] and organochlorine pesticides), polybrominated diphenyl ethers, per- and polyfluoroalkyl substances (PFAS), metals, phthalate metabolites, phenols, organophosphate pesticide metabolites, and cotinine). Lifestyle factors included smoking habits, physical activity, allergens, sleep, and socioeconomic status. Details on the correlations between assessed exposures has been previously documented 19 Additional information regarding the type of assessed exposures can be found in the Supplementary Table S2. Informed consent was available for the use of the data as described in this manuscript and ethics approval for such use was available.

The environmental exposures considered in this analysis encompassed outdoor factors, chemical constituents, and lifestyle determinants. Outdoor exposures were categorized into built environment parameters, meteorological conditions, natural surroundings, road traffic influences, and noise levels. Chemical exposures included a broad spectrum of substances such as organochlorine compounds (including polychlorinated biphenyls [PCBs] and organochlorine pesticides), polybrominated diphenyl ethers, per- and polyfluoroalkyl substances (PFAS), various metals, phthalate metabolites, phenols, organophosphate pesticide metabolites, and cotinine. Lifestyle factors examined encompassed smoking habits, levels of physical activity, exposure to allergens, sleep patterns, and socioeconomic status. For further details on the correlations between these assessed exposures, readers are referred to previous documentation[38]. Additional information pertaining to the types of exposures assessed can be found in **Table S2**. Informed consent was obtained for the usage of the data as outlined in this manuscript, and ethical approval for such usage was obtained. The outcome variable was represented as a binary variable indicating the diagnosis of hypertension, in accordance with the 2004 reference charts from the National High Blood Pressure Education Program. The outcome has been previously investigated in this population, and the main findings of the research can be found in[27,28].

In total, 1143 healthy and 155 children diag nosed with hypertension are included in the analysis. 77 early exposures are included as the features, together with inferred 13 covariates. Prior to analysis, all variables included in the dataset underwent appropriate preprocessing steps, which involved the removal of outliers, normalization and standardization of features, and imputation of missing values, as detailed in previous literature[18].

*Covariate inference*

Exposure data was available for both the prenatal and postnatal periods, necessitating the execution of separate models for each time point. To account for potential confounding factors, a Directed Acyclic Graph (depicted in **Figure S1**) was utilized, identifying 13 variables as potential confounders. These variables included factors from both the prenatal and postnatal periods.

The confounding variables encompassed the cohort of recruitment, maternal age (continuous in years), maternal educational level (categorized as low, middle, or high), self-reported maternal pre-pregnancy body mass index (continuous in kg/m2), parity (classified as nulliparous, primiparous, or multiparous), native status (indicating whether the child's family is native to the country of recruitment), as well as the child's age, height, and sex. Cohort of recruitment was represented by six dummy variables, one for each population involved, with only five of them being included as confounders in the analysis. This approach ensured proper adjustment for potential biases arising from cohort differences.

## Introduction of DataSHIELD

Whilst several tools and platforms have recently become available in federated learning[39], the so-called 'DataSHIELD' infrastructure has gained considerable attention in the epidemiological and biomedical fields[40]. DataSHIELD is an R-based software solution for individual level federated analysis using disclosure-preventing methods to ensure ethico-legal compliance[20,21]. Principal features of DataSHIELD involve: I) a client-server architecture, II) a bunch of builtin and community-developed federated packages[41] for, e.g., federated visualization as well as omics analysis, and III) tailored multi-layer disclosure controls (to ensure that the analyst cannot see, copy or extract individual level data held by individual studies). The key advantages of DataSHIELD include its open-source nature, that is written in R, and is licensed under the GPL, thus facilitating downstream analyses within a single pipeline by interacting with other programming languages (e.g., Python) and with other R or Bioconductor packages.

DataSHIELD is currently used within a number of H2020 and Horizon Europe projects (ref LifeCycle, ATHLETE). Whilst a wide range of statistical methods have been implemented within DataSHIELD, there is currently a lack of state-of-the-art machine learning (ML) approaches commonly used in biomedicine and epidemiology[10,19]. Machine learning techniques include methods for variable selection, aimed at identifying the 'best' subset of features for predicting a given study outcome or phenotype. These techniques can increase the predictive capacity and interpretability of models while tackling overfitting when applied to high-dimensional data, such as those often found in the exposome and omics. A key method currently previously unavailable in DataSHIELD is Lasso regression – a regularization technique similar to ordinary least squares, except that it imposes a shrinkage penalty on the magnitude of the estimated coefficients[22,24]. Lasso is among the most straightforward and frequently employed feature-selection techniques[23], specially for the analysis of complex epidemiological datasets[42]. In the current paper, we describe some of the challenges of implementing LASSO in a federated setting, and present 'dsLassoCov', an implementation within DataSHIELD of the Lasso feature-selection method.

1. Sun, S., Wang, C., Ding, H. & Zou, Q. Machine learning and its applications in plant molecular studies. *Briefings in functional genomics* **19**, 40-48 (2020).
2. Watson, D.S. *et al.* Clinical applications of machine learning algorithms: beyond the black box. *Bmj* **364**(2019).
3. Asiimwe, R. *et al.* From biobank and data silos into a data commons: convergence to support translational medicine. *Journal of Translational Medicine* **19**, 1-13 (2021).
4. Voigt, P. & Von dem Bussche, A. The eu general data protection regulation (gdpr). *A Practical Guide, 1st Ed., Cham: Springer International Publishing* **10**, 10-5555 (2017).
5. Decherchi, S., Pedrini, E., Mordenti, M., Cavalli, A. & Sangiorgi, L. Opportunities and challenges for machine learning in rare diseases. *Frontiers in medicine* **8**, 747612 (2021).
6. Mammen, P.M. Federated learning: Opportunities and challenges. *arXiv preprint arXiv:2101.05428* (2021).
7. Doiron, D. *et al.* Data harmonization and federated analysis of population-based studies: the BioSHaRE project. *Emerg Themes Epidemiol* **10**, 12 (2013).
8. Wilson, R.C. *et al.* DataSHIELD – New Directions and Dimensions. *Data Science Journal* **16**(2017).
9. Nasirigerdeh, R. *et al.* sPLINK: A Federated, Privacy-Preserving Tool as a Robust Alternative to Meta-Analysis in Genome-Wide Association Studies. *BioRxiv* (2020).
10. Marcon, Y. *et al.* Orchestrating privacy-protected big data analyses of data from different resources with R and DataSHIELD. *PLoS Comput Biol* **17**, e1008880 (2021).
11. Wang, J. & Ma, F. Federated learning for rare disease detection: a survey. *Rare Disease and Orphan Drugs Journal* **16**(2023).
12. Li, T. *et al.* Federated optimization in heterogeneous networks. *Proceedings of Machine learning and systems* **2**, 429-450 (2020).
13. Leek, J.T. *et al.* Tackling the widespread and critical impact of batch effects in high-throughput data. *Nature Reviews Genetics* **11**, 733-739 (2010).
14. Jager, K.J., Zoccali, C., Macleod, A. & Dekker, F.W. Confounding: what it is and how to deal with it. *Kidney Int* **73**, 256-60 (2008).
15. Snoek, L., Miletić, S. & Scholte, H.S. How to control for confounds in decoding analyses of neuroimaging data. *Neuroimage* **184**, 741-760 (2019).
16. Brenner, H. & Blettner, M. Controlling for continuous confounders in epidemiologic research. *Epidemiology* **8**, 429-434 (1997).
17. Frisch, R. & Waugh, F.V. Partial time regressions as compared with individual trends. *Econometrica: Journal of the Econometric Society*, 387-401 (1933).
18. Maitre, L. *et al.* Human Early Life Exposome (HELIX) study: a European population-based exposome cohort. *BMJ Open* **8**, e021311 (2018).
19. Cao, H. *et al.* dsMTL - a computational framework for privacy-preserving, distributed multi-task machine learning. *Bioinformatics* (2022).
20. Gaye, A. *et al.* DataSHIELD: taking the analysis to the data, not the data to the analysis. *International journal of epidemiology* **43**, 1929-1944 (2014).
21. Jaddoe, V.W.V. *et al.* The LifeCycle Project-EU Child Cohort Network: a federated analysis infrastructure and harmonized data of more than 250,000 children and parents. *Eur J Epidemiol* **35**, 709-724 (2020).
22. Tibshirani, R. Regression shrinkage and selection via the lasso. *Journal of the Royal Statistical Society. Series B (Methodological)*, 267-288 (1996).
23. Zou, H. & Hastie, T. Regularization and variable selection via the elastic net. *Journal of the Royal Statistical Society: Series B (Statistical Methodology)* **67**, 301-320 (2005).

24. Tibshirani, R. Regression shrinkage and selection via the lasso: a retrospective. *Journal of the Royal Statistical Society: Series B (Statistical Methodology)* **73**, 273-282 (2011).
25. Chernozhukov, V. *et al.* Double/debiased machine learning for treatment and structural parameters. (Oxford University Press Oxford, UK, 2018).
26. Urminsky, O., Hansen, C. & Chernozhukov, V. Using Double-Lasso Regression for Principled Variable Selection. *SSRN Electronic Journal* (2016).
27. Warembourg, C. *et al.* Urban environment during early-life and blood pressure in young children. *Environ Int* **146**, 106174 (2021).
28. Warembourg, C. *et al.* Early-Life Environmental Exposures and Blood Pressure in Children. *J Am Coll Cardiol* **74**, 1317-1328 (2019).
29. Vrijheid, M. *et al.* Advancing tools for human early lifecourse exposome research and translation (ATHLETE): Project overview. *Environ Epidemiol* **5**, e166 (2021).
30. Cao, H., Chen, J., Meyer-Lindenberg, A. & Schwarz, E. A polygenic score for schizophrenia predicts glycemic control. *Transl Psychiatry* **7**, 1295 (2017).
31. Elze, M.C. *et al.* Comparison of propensity score methods and covariate adjustment: evaluation in 4 cardiovascular studies. *Journal of the American College of Cardiology* **69**, 345-357 (2017).
32. Nesterov, Y. Gradient methods for minimizing composite functions. *Mathematical Programming* **140**, 125-161 (2012).
33. Beck, A. & Teboulle, M. A fast iterative shrinkage-thresholding algorithm for linear inverse problems. *SIAM journal on imaging sciences* **2**, 183-202 (2009).
34. Liu, J. & Jieping, Y. Efficient L1/Lq Norm Regularization.
35. Peng, Z., Xu, Y., Yan, M. & Yin, W. ARock: An Algorithmic Framework for Asynchronous Parallel Coordinate Updates. *SIAM Journal on Scientific Computing* **38**, A2851-A2879 (2016).
36. Breiman, L. Random forests. *Machine learning* **45**, 5-32 (2001).
37. Chang, C.-C. & Lin, C.-J. LIBSVM: a library for support vector machines. *ACM transactions on intelligent systems and technology (TIST)* **2**, 1-27 (2011).
38. Tamayo-Uria, I. *et al.* The early-life exposome: description and patterns in six European countries. *Environment international* **123**, 189-200 (2019).
39. Mullie, L. *et al.* CODA: an open-source platform for federated analysis and machine learning on distributed healthcare data. *Journal of the American Medical Informatics Association* **31**, 651-665 (2024).
40. Wolfson, M. *et al.* DataSHIELD: resolving a conflict in contemporary bioscience—performing a pooled analysis of individual-level data without sharing the data. *International journal of epidemiology* **39**, 1372-1382 (2010).
41. community, D. Packages developed in DataSHIELD.
42. Santos, S. *et al.* Applying the exposome concept in birth cohort research: a review of statistical approaches. *European journal of epidemiology* **35**, 193-204 (2020).

## Supplementary Tables

## Table S1 – Description of the study population

| Variable name | N (%) | Min | Q1 | Median | Q3 | Max |
|---|---|---|---|---|---|---|
| Maternal age at birth | 1298 (100 %) | 16 | 27.3 | 31 | 34. | 43.5 |
| Pre-pregnancy BMI | 1298 (100 %) | 15.9 | 21.3 | 23.8 | 27.1 | 51.4 |
| Child age at the follow up | 1298 (100 %) | 5.4 | 6.5 | 8 | 8.9 | 12.1 |
| SBP (mmHg) | 1298 (100 %) | 70.5 | 92 | 98.5 | 106.5 | 159 |
| DBP (mmHg) | 1298 (100 %) | 37 | 52 | 57 | 62.5 | 118.5 |
| | | | | **Category** | **n (%)** | |
| Hypertension | 1298 (100 %) | | | Control | 1143 (88.1 %) | |
| | | | | Case | 155 (11.9 %) | |
| Cohort | 1298 (100 %) | | | BiB | 202 (15.6%) | |
| | | | | EDEN | 198 (15.2%) | |
| | | | | INMA | 223 (17.2%) | |
| | | | | KANC | 204 (15.7%) | |
| | | | | MoBA | 272 (21%) | |
| | | | | RHEA | 199 (15.3%) | |
| Highest maternal education | 1298 (100 %) | | | Primary | 180 (13.9%) | |
| | | | | Secondary | 444 (34.2%) | |
| | | | | Higher | 674 (51.9%) | |
| Child sex | 1298 (100 %) | | | Girl | 590 (45.5%) | |
| | | | | Boy | 708 (54.5%) | |

Population: study population from the HELIX subcohort, n=1298 children.
Acronyms: DBP, Diastolic Blood Pressure, SBP: Systolic Blood Pressure, BiB: Born in Bradford, EDEN: Étude des Déterminants pré et postnatals du développement et de la santé de l'Enfant, INMA: Infancia y Medio Ambiente, IOTF: International Obesity Task Force, KANC: Kaunus Cohort, MoBa: The Norwegian Mother, Father and Child Cohort Study, RHEA: Mother-Child Cohort in Crete.

# Table S2 – List of all exposures measured and studied

| Type of exposure | Exposures | Number of exposures Childhood |
|---|---|---|
| **URBAN EXPOSOME** | | |
| Outdoor air pollution | $NO_2$, $PM_{10}$, $PM_{2.5}$, $PM_{2.5}$ absorbance | 4 |
| Indoor air pollution | $NO_2$, $PM_{2.5}$, PM absorbance, Benzene | 4 |
| Meteorology | Temperature, UV | 2 |
| Surrounding natural spaces | NDVI | 1 |
| Built environment | Accessibility, population density, building density, street connectivity, accessibility, facility richness, walkability, land use index | 8 |
| Road traffic and noise | Traffic load on all roads and nearest road, traffic density on nearest road, inverse distance to nearest road. | 3 |
| Total of urban exposures for each period | | 22 |
| **LIFESTYLE** | | |
| Diet | KIDMED score | 1 |
| Physical activity | Moderate/vigorous activity, sedentary time | 2 |
| Socio-economic & others | House crowding, self-perceived stress score | 2 |
| Total of lifestyle exposures for each group | | 5 |
| **CHEMICAL EXPOSOME** | | |
| Organochlorine pesticides (OCs) | PCB, DDE, DDT, HCB | 8 |
| Perfluoroalkyl substances (PFASs) | PFHxS, PFOS, PFOA, PFNA, PFUnDA | 5 |
| Brominated compounds (PBDEs) | PBDE 47, 153 | 2 |
| Metal and essential elements | As, Hg, Cd, Pb, Cs, Cu, Mn, Co, Mo, Se, Tl | 14 |
| Phthalates | MECPP, MEHHP, MEHP, MEOHP, MEP, MiBP, MnBP, MBzP, DEHP [a], oxo-MiNP and oh-MiNP | 10 |
| Phenols | MEPA, ETPA, PRPA, BUPA, BPA, OXBE, TRCS | 7 |
| Organophosphate (OP) metabolites | DMP, DMTP, DEP, DETP, DMDTP | 4 |
| Total of chemical exposures for each group | | 50 |
| **Total of exposures** | | **77** |

a DEHP: molar sum of MEHP, MEHHP, MEOHP and MECPP.

Table S3. Estimated Odd ratios in the real HELIX dataset (Model: HT ~ exposome + confounders).

| Feature | OR (dsLassoCov) | OR (local-dsLassoCov) | Group |
|---|---|---|---|
| Cohort_KANC | **0.016** | **0.012** | Confounder |
| Cohort_RHEA | **0.019** | **0.012** | Confounder |
| Cohort_BIB | **0.111** | **0.119** | Confounder |
| Cohort_MOBA | **0.136** | **0.139** | Confounder |
| Cohort_INMA | **0.311** | **0.370** | Confounder |
| Child age (days) | **0.429** | **0.414** | Confounder |
| Maternal education | **0.795** | **0.802** | Confounder |
| Country native | **0.866** | **0.877** | Confounder |
| MBzP | 0.910 | 0.935 | Phthalates |
| Accessibility (bus stops 300m) | 0.920 | 0.869 | Built Environment |
| Land use (300m) | 0.928 | 0.941 | Built Environment |
| Arsenic | 0.939 | 0.950 | Metals |
| Parity | **0.969** | **0.986** | Confounder |
| Cobalt | 0.969 | 0.000 | Metals |
| PBDE 153 | **0.970** | **0.989** | PBDEs |
| Lead | 0.978 | 0.000 | Metals |
| DMTP | 0.985 | 0.000 | OP Pesticides |
| MEHP | 0.994 | 0.000 | Phthalates |
| Sex of child | **0.996** | **1.003** | Confounder |
| MnBP | 0.997 | 0.000 | Phthalates |
| PCB 118 | 0.998 | 0.000 | OCs |
| PCB 138 | 0.998 | 0.000 | OCs |
| Inverse distance to nearest road (home) | 0.999 | 0.000 | Traffic |

| Feature | Odds Ratio | | Category |
|---|---|---|---|
| DDE | 0.999 | 0.000 | OCs |
| MEP | 0.999 | 0.000 | Phthalates |
| PCB 180 | 0.999 | 0.000 | OCs |
| Population density | 1.0003 | 0.000 | Built Environment |
| OXBE | 1.001 | 0.000 | Phenols |
| Moderate and vigorous PA | 1.003 | 0.000 | Lifestyle |
| Sedentary behaviour | 1.004 | 0.000 | Lifestyle |
| UV - Vit.D (month) | 1.004 | 0.000 | Meteorological |
| Manganese | 1.007 | 0.000 | Metals |
| oxo-MiNP | 1.014 | 0.000 | Phthalates |
| Indoor PM2.5 | 1.014 | 0.000 | Indoor air |
| PFNA | 1.015 | 0.000 | PFASs |
| PMabsorbance (year) | 1.021 | 0.000 | Air Pollution |
| Height (m) | **1.029** | **1.013** | Confounder |
| NO2 (year) | 1.032 | 0.000 | Air Pollution |
| <u>Built density (300m)</u> | **1.045** | **1.030** | Built Environment |
| <u>BPA</u> | **1.061** | **1.042** | Phenols |
| <u>Perceived stress score</u> | **1.077** | **1.068** | Others |
| <u>Magnesium (Mg) in child</u> | **1.104** | **1.082** | Essential minerals |
| <u>PFOA</u> | **1.107** | **1.082** | PFASs |
| Maternal age (years) | **1.112** | **1.090** | Confounder |
| Maternal pre-pregnancy bmi | **1.245** | **1.249** | Confounder |

Confounders were forced to retain in the final output model in both dslassoCov and local-dslassoCov. Features are in increasing order according to the estimated odd ratio. Odd ratios in bold correspond to features selected in both dsLassoCov and local-dslassoCov models. Underlined features correspond to exposures selected by the two models that are not confounders.

Supplementary Figures

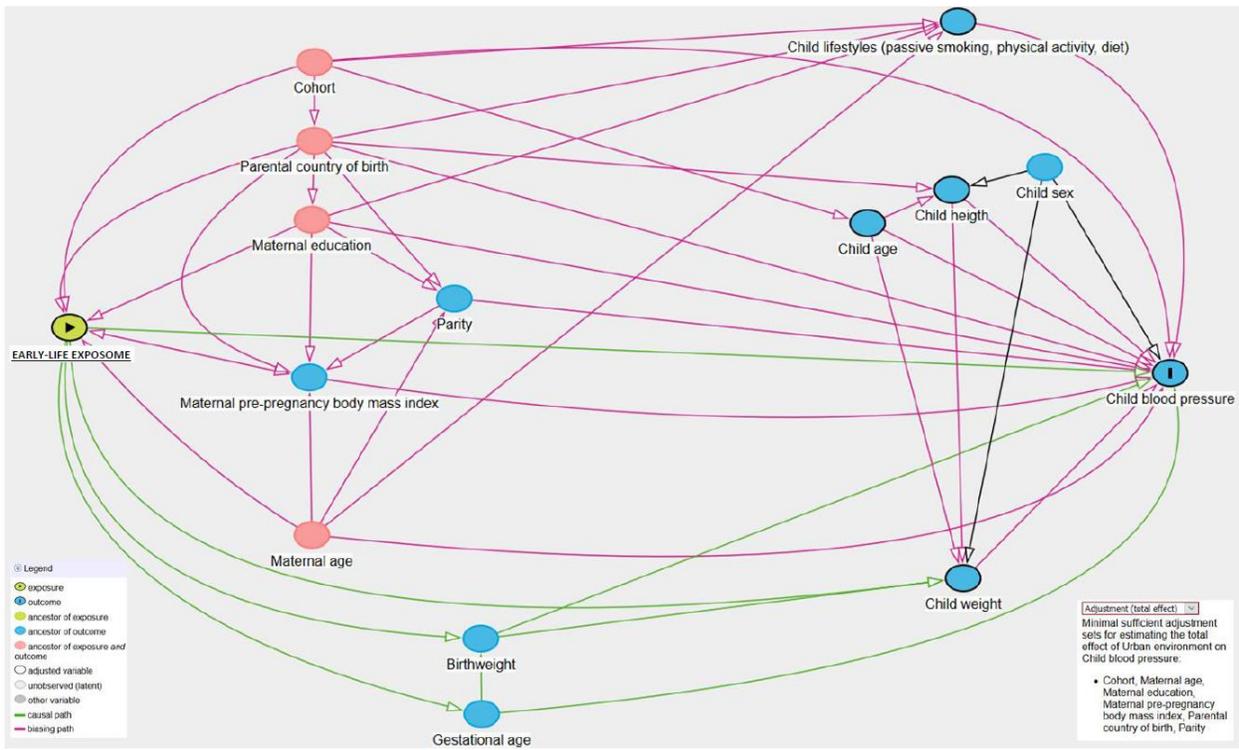

Figure S1, Direct Acyclic Graph design to identified confounder in the association between the early-life exposome and child blood pressure.